# From Nuclear Reactions to High–Frequency Trading: an R–function Approach

Frank W K Firk

The Henry Koerner Center for Emeritus Faculty,

Yale University, New Haven, CT  06520

**Abstract.**  The R-function theory of Thomas (1955) – a practical version of the general R-matrix theory of Wigner and Eisenbud (1947) – is used to calculate the fine, intermediate and gross structure observed in the energy-dependence of nuclear reactions, and in the time-dependence of the Dow Jones Industrial Average, a key economic index.  In these two disparate fields, the three basic structures are characterized by the values of the fundamental "strength function", $<\Gamma>/<D>$ where $<\Gamma>$ is the average width (lifetime) of the underlying states, and $<D>$ is the average spacing between adjacent states.  It is proposed that the values of $<\Gamma>/<D>$ for the fine and intermediate structure of the index, determined in the first hour of trading on a given day, provide valuable information concerning the likely performance of the index for the remainder of the trading day.  The universality of fluctuations observed in resonating systems is discussed in terms of the statistical properties of complex bivariates.

NUCLEAR REACTIONS.  The R-function theory of Thomas is used to model neutron inelastic scattering to a definite state and to model the fine, intermediate, and gross structure observed in the Dow Jones Industrial Average on a typical trading day.

# 1. INTRODUCTION

An important parameter in systems that exhibit time-dependent structures (resonances or fluctuations of a general nature), both classical and quantum, is the ratio, average width (lifetime) of the states/ average spacing of the states, denoted by $\langle\Gamma\rangle/\langle D\rangle$. If $\langle\Gamma\rangle/\langle D\rangle \ll 1$, the states are clearly separated, and the parameters that characterize individual states can be determined with an exact theory (nuclear resonance theories are discussed in the standard work of Lynn, 1968). As the complexity of the system increases, a region is reached in which $\langle\Gamma\rangle/\langle D\rangle > 1$, and new challenges for both observers and analysts are encountered. This is due to the fact that, at a given energy or time, many overlapping local and distant states contribute to the response of the system, and all the states must be treated coherently. If the systems are studied as a function of energy, time, or frequency, constructive and destructive interference effects generate "fluctuations" in the observed spectra. The origin of fluctuations differs in a fundamental way from the origin of "collective" states. The exact theory of strongly overlapping states is, in general, intractable; statistical arguments must then be used.

A striking similarity is observed between the *forms* of the fine, intermediate, and gross structure of photonuclear states, studied as a function of energy in medium-mass nuclei, and the fine, intermediate, and gross structure observed in the fluctuations in a typical economic index over a period of time.   Examples are shown in Figure 1.

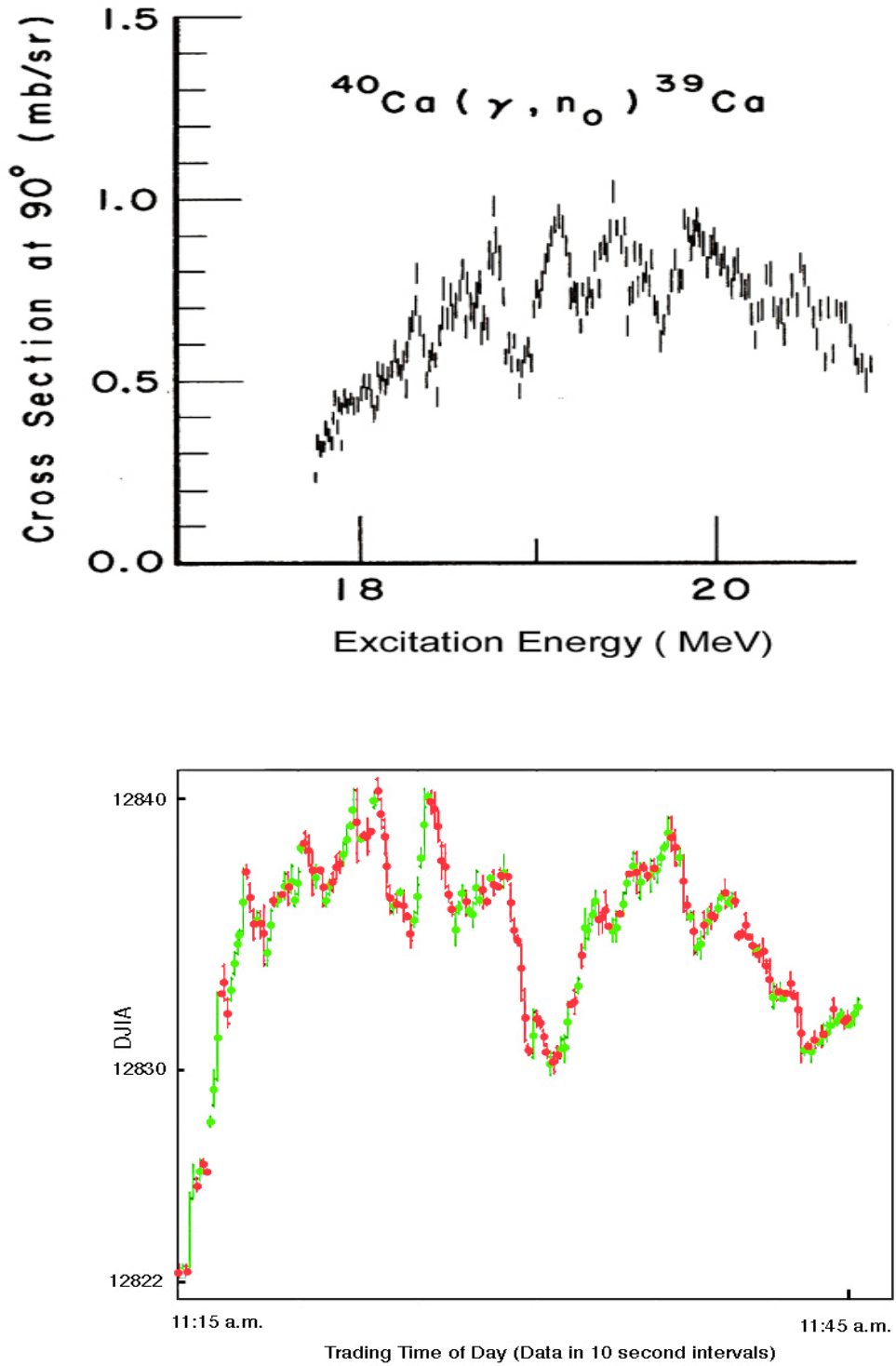

Figure 1. *The cross section for the reaction $^{40}$ Ca(γ, n$_o$) $^{39}$Ca (Wu, Firk, and Phillips, 1970) in the region of the giant resonance, and the observed DJIA in a 30-minute interval, displayed in 10-second intervals*.

In nuclei, the gross, intermediate and fine structures are attributed to the evolution of nuclear reactions from the underlying "single-particle" states to "few-particle-few-hole" states to "many-particle-many-hole" compound states, respectively (see Bohr and Mottelson, 1975). In the fluctuations of an economic index, the three structural forms are associated with long, medium, and short-range time-correlations in the complex trading options that take place among all elements that contribute to the index. *The fine structure is a characteristic feature of High-Frequency Trading.*

The market strategy of buying low and selling high is a reflection of a lack of a sophisticated model of the *structure* associated with a typical economic index. It is necessary to understand the *mathematical form* of the time-dependence of the fluctuations; the R-function theory described here provides an analytical means for studying the individual and the average properties of the states. The theory includes, explicitly, the all-important "lifetime" of each state,

In the nuclear case, the origin of the intermediate structure is not always clear. The observed form can result not only from few-particle-few-hole states but also from Ericson fluctuations (Ericson, 1960, 1963). In the model of the DJIA presented here, the origin of the intermediate structure is not of primary importance; it is the mathematical method used to describe the structural form that matters.

In a seminal paper, Thomas (1955), developed the exact theory of resonant quantum states (Wigner and Eisenbud (1947)) to include cases in which $<\Gamma>/<D> > 1$. Using arguments that were of practical importance in the study of nuclear resonances, he developed the theory needed to analyze observed spectra. His method has formed the basis of a great deal of the subsequent work in the field of nuclear spectroscopy (Vogt (1958), Reich and Moore (1958), Firk, Lynn and Moxon (1963)). Thomas also discussed the forms of nuclear cross sections when averaged over large energy intervals.

A major development in dealing with energy-averaged nuclear cross sections was made by Ericson (1960,1963). He showed that in regions in which $<\Gamma>/<D> > 1$, fluctuations naturally occur. Ericson averaged the scattering amplitude given by Feshbach (1962), and Moldauer (1964) refined the method by using the statistical arguments of Porter and Thomas (1956) to deal with the average properties of the widths (lifetimes) of the overlapping resonances.

Thomas' R–function theory is developed to describe the fine, intermediate, and gross structures observed in both inelastic neutron scattering, and the prices of stocks reported on a typical trading day.

## 2. ON THE UNIVERSALITY OF FLUCTUATIONS

Universality in physical systems is frequently discussed in terms of a common underlying mathematical structure. We are therefore led to ask "is there a mathematical structure associated with universal fluctuations, and, if so what is it?" A clue to answering this question is obtained by noting that the amplitudes of resonating systems have the complex forms

$f = a + ib$ .

(Recall that, a lightly damped linear oscillator with a resonant frequency $\omega_R$, has an amplitude, A, when driven with a driving frequency $\omega$, given by

$$A = (\Gamma/2)/[(\omega - \omega_R) + i(\Gamma/2)],$$

and an intensity

$$I(\omega) = AA^* = (\Gamma/2)^2/[(\omega - \omega_R)^2 + (\Gamma/2)^2], \text{ a Lorentzian,}$$

where $\Gamma$ is the width of the resonance).

We therefore take the following approach: let a set of complex numbers

$$z = x + iy$$

be statistically distributed, and let their mean values and mean complex squares be equal to zero. Fluctuations of z about the mean value occur in a random way. We have

$$\langle z \rangle = \langle x \rangle + i\langle y \rangle = 0$$

and therefore

$$\langle x \rangle = \langle y \rangle = 0.$$

Also,

$$\langle z^2 \rangle = \langle x^2 \rangle + 2i\langle xy \rangle - \langle y^2 \rangle = 0$$

and therefore

$$\langle x^2 \rangle = \langle y^2 \rangle = s^2 \text{ (say)},$$
$$\langle xy \rangle = 0,$$

and,

$$\langle |z|^2 \rangle = \langle x^2 \rangle + \langle y^2 \rangle \neq 0.$$

The probability density function of a bivariate distribution is

$$P(x, y) = (1/2\pi s_x s_y(1 - \rho^2))(\exp\{-G\})$$

where

$$G = (1/2(1 - \rho^2))[(x - \mu_x)^2/s_x^2 - 2\rho(x - \mu_x)(y - \mu_y)/s_x s_y + (y - \mu_y)^2/s_y^2,$$

ρ is the correlation coefficient, $\mu_x$, $\mu_y$ are the means, and $s_x$, $s_y$ are the standard deviations of x and y, respectively.

If the variables x and y have independent normal distributions, and they have the same variance, $s^2$, then their bivariate probability density distribution is

$$P(x, y) = [1/(2\pi s^2)]\exp\{-(x^2 + y^2)/2s^2\}.$$

$|z|^2$, given by the sum of the squares of two uncorrelated variables with the same variance, $s^2$, has a probability distribution that belongs to the $\chi^2$ – family with two degrees of freedom, an exponential distribution:

$$P(w) = [\exp\{-w/2\}]/2$$

where

$$w = |z|^2/<|z|^2>.$$

Significant fluctuations of w about the mean value are predicted.

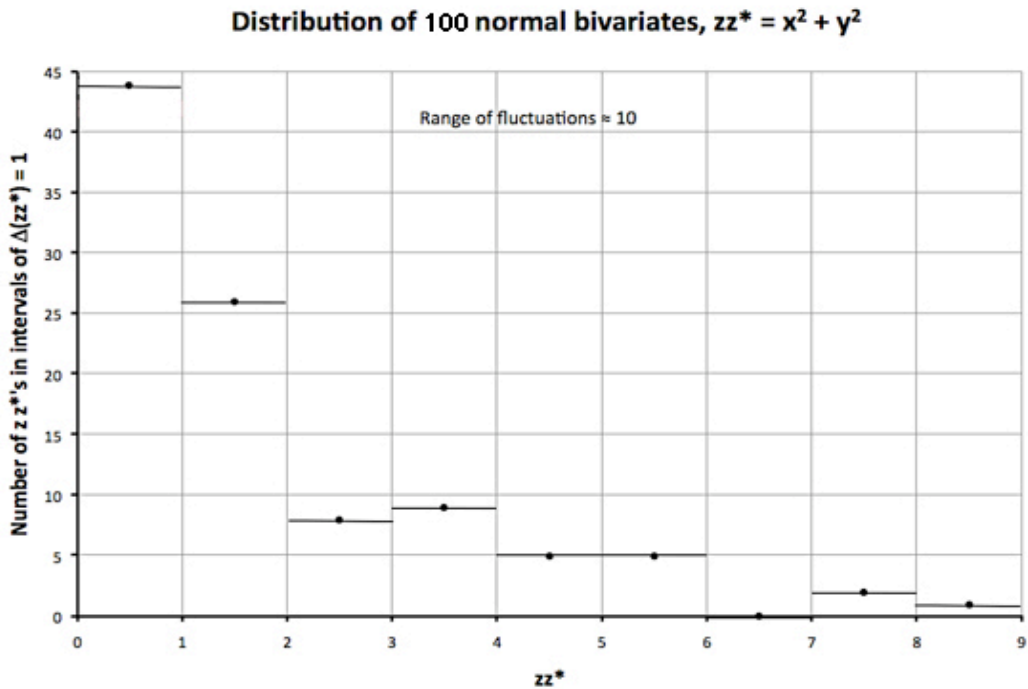

Figure 2. *The calculated distribution of 100 independent, randomly chosen, normal bivariates.*

In this set, containing 100 members, the ratio of largest-to-least value of w is ~10.

*The above statistical approach can be applied to all systems in which the complex values (amplitudes) of z generate resonant-like structures, and their real and imaginary parts belong to some probability distribution functions.*

Ericson considered an autocorrelation function, $C(\Delta n)$, defined as

$C(\Delta n) = <[z(n + \Delta n) - <z>][z(n) - <z>]>/<z>^2$

where

$<z> = (1/N)\sum_{n=1,N} z(n)$.

The normalized variance is given when $\Delta n = 0$:

$C(0) = \{<z^2> - <z>^2\}/<z>^2$.

Amplitude correlations involve the autocorrelation function, $A(\Delta n)$:

$A(\Delta n) = \{<z(n + \Delta n)z^*(n)> - |<z>|^2\}/<z>$

If the correlations have normal distributions with zero mean then

$C(\Delta E) = |A(\Delta E)|^2$.

If the scattering amplitude $z \to f(E)$, in the region of a resonance at an energy $E_r$, with total width $\Gamma$, has the form

$f(E) \sim 1/[(E - E_r) + i\Gamma/2]$

then the amplitude autocorrelation function can be written

$A(\Delta E) = 1/[1 - i(\Delta E/\Gamma)]$.

The energy autocorrelation function then becomes

$C(\Delta E) = |A(\Delta E)|^2 = \Gamma^2/[(\Delta E)^2 + \Gamma^2]$.

This is a key result in Ericson's theory.

At sufficiently high excitation energy, the average width $<\Gamma>$ becomes greater than the average spacing $<D>$ between the resonances.

The total width, made up of many partial widths, then becomes essentially constant. At a given energy, the contributions from the many overlapping resonances must be treated coherently. The average width <Γ> is therefore referred to as the "coherence energy".

In practice, it is often difficult to study a set of resonances of the same spin and parity, and to know all the possible decay modes; the correlations are then dampened. Also, the effects of finite resolution of the spectrometer used to measure the cross section can artificially reduce the correlation effects. If possible contributions to the cross section from non-compound nuclear effects are included, the problem of analysis becomes even more challenging.

## 3. A MICROSCOPIC APPROACH TO FLUCTUATIONS

Thomas' theory (1955) provides a method of dealing with many overlapping resonances (<Γ>/<D> > 1). A problem of a general nature concerns the inelastic scattering of a particle from highly excited states of a many-body system to a definite final state. The inelastic scattering is accompanied by elastic scattering, and by transitions to many alternative channels. Here, a specific problem is considered in which inelastic scattering of a neutron by an even-even nucleus to a definite state is considered (see Firk (2010) for a detailed discussion). Elastic scattering is included explicitly, and inelastic scattering and radiative capture, to all other allowed states, are treated in an average way. Each state, λ, with a spin and parity $J^\pi$ is characterized by an energy $E_\lambda$ and a total width $\Gamma_\lambda$. An incident neutron interacts with a heavy nucleus to form a state that decays into many different channels. The two main channels are inelastic scattering to a definite state (width $\Gamma_{n'}$), and elastic scattering (width $\Gamma_n$). All the subsidiary channels involve inelastic scattering (width $\Gamma_{n''i}$) and radiative capture (width $\Gamma_{\gamma i}$).

3.1. *The Thomas approximation*

The **R**-matrix has the energy-dependent form (Lane and Thomas, 1958)

$$\mathbf{R}(E) = \sum_\lambda \gamma_{\lambda c'}\gamma_{\lambda c''}/(E_\lambda - E)$$

where the sum is over *all* levels $\lambda$ of energy $E_\lambda$, and the $\gamma_{\lambda c}$'s are the reduced width amplitudes associated with the channels c', c". If the signs of the amplitudes are sufficiently random to ensure that the non-diagonal elements of **R** are small compared with the diagonal elements, Thomas showed that the resulting collision matrix **U** can be written

$$\mathbf{U} = \sum_\lambda (a_{\lambda c'}\, a_{\lambda c''})/(E_\lambda - E - i\Gamma_\lambda/2),\ c' \neq c'',$$

(an "amplitude" in the spirit of the earlier discussion) even for *overlapping states*. The quantities, $a_{\lambda c}$'s are given by

$$a_{\lambda c} = \gamma_{\lambda c}\sqrt{(2P_c)},$$

where $P_c$ is the penetration factor, and the width is

$$\Gamma_{\lambda c} = a_{\lambda c}^2.$$

The width that occurs in the denominator is

$$\Gamma_\lambda = \sum_c \Gamma_{\lambda c},$$

in which

$$\Gamma_{\lambda c} = |a_{\lambda c}/(1 - iP_c(R_c^\infty + i\pi\rho<\gamma_{\lambda c}>^2))|^2,$$

$R^\infty$ is the effect of all states outside the range of interest, $\rho$ is the density of states, and <...> denoted the average value in the region of E.

The cross section for the reaction c' –> c" is

$$\sigma_{c'c''} = (\pi/k_{c'})^2 \sum_{c',c''} |U_{c',c''}|^2$$

where $k_{c'}$ is the wave number of the relative motion of the two particles in the incident channel. (The spin weighting factor has been put equal to unity).

In the present case that involves two main channels and many subsidiary channels, Thomas (1955) showed that a *reduced* form of the **R**-matrix is valid; it is

$$R(E) = \sum_\lambda (\gamma_{\lambda c'} \gamma_{\lambda c''})/(E_\lambda - E - i\Gamma_\lambda^e/2), \text{ the } R\text{-function},$$

where $\Gamma_\lambda^e$ is a suitable average of the widths of all the subsidiary, or *eliminated* channels. Here, the eliminated channels are all channels except the incident channel, and the inelastic scattering channel to the definite state. *The reduced **R**-matrix is valid if the means of the partial widths for the eliminated channels are less than the spacings, and their reduced width amplitudes are random in sign.* It is seen that the reduced **R**-matrix can be obtained from the traditional **R**-matrix by evaluating it at the complex energy $E = E + i\Gamma_\lambda^e/2$.

3.2 *Cross section for inelastic neutron scattering to a definite state*

The cross section for the inelastic scattering of a neutron to a definite state in the presence of elastic, all other inelastic channels, and radiative capture is calculated using Thomas' R-function:

$$\sigma_{n,n'}(k_n^2/4\pi) = \left| \frac{\sum_\lambda (\Gamma_{\lambda n}/2)^{1/2}(\Gamma_{\lambda n'}/2)^{1/2}/f_\lambda(E)}{[1 - i\sum_\lambda(\Gamma_{\lambda n}/2)/f_\lambda(E)][1 - i\sum_\lambda(\Gamma_{\lambda n'}/2)/f_\lambda(E)] + [\sum_\lambda(\Gamma_{\lambda n}/2)^{1/2}(\Gamma_{\lambda n'}/2)^{1/2}/f_\lambda(E)]^2} \right|^2$$

where

$$f_\lambda(E) = E_\lambda - E - i\Gamma_\lambda^e/2,$$

and the sums are over *all* states $\lambda$ of the same spin and parity. The width of the eliminated channels is

$$\Gamma_\lambda^e = \sum_{n''}\Gamma_{\lambda n''} + \sum_i \Gamma_{\lambda \gamma i} = \Gamma_\lambda - (\Gamma_{\lambda n} + \Gamma_{\lambda n'}),$$

where $\Gamma_\lambda$ is the total width, $\Gamma_{\lambda n}$ is the elastic scattering width, $\Gamma_{\lambda n'}$ is the inelastic scattering width to the definite state, $\Gamma_{\lambda n''}$ is an inelastic scattering width to an eliminated state, and $\Gamma_{\lambda \gamma i}$ is a partial radiation width to an eliminated state. The neutron wave number associated with the incident channel is $k_n$. The cross section $\sigma_{nn'}$ is calculated for up to 1000 interfering states at up to 10000 energies.

### 3.3 *Onset of fluctuations*

The calculated inelastic neutron scattering cross section for two values of the strength function, $<\Gamma>/<D> = 0.4$, and 7 is shown:

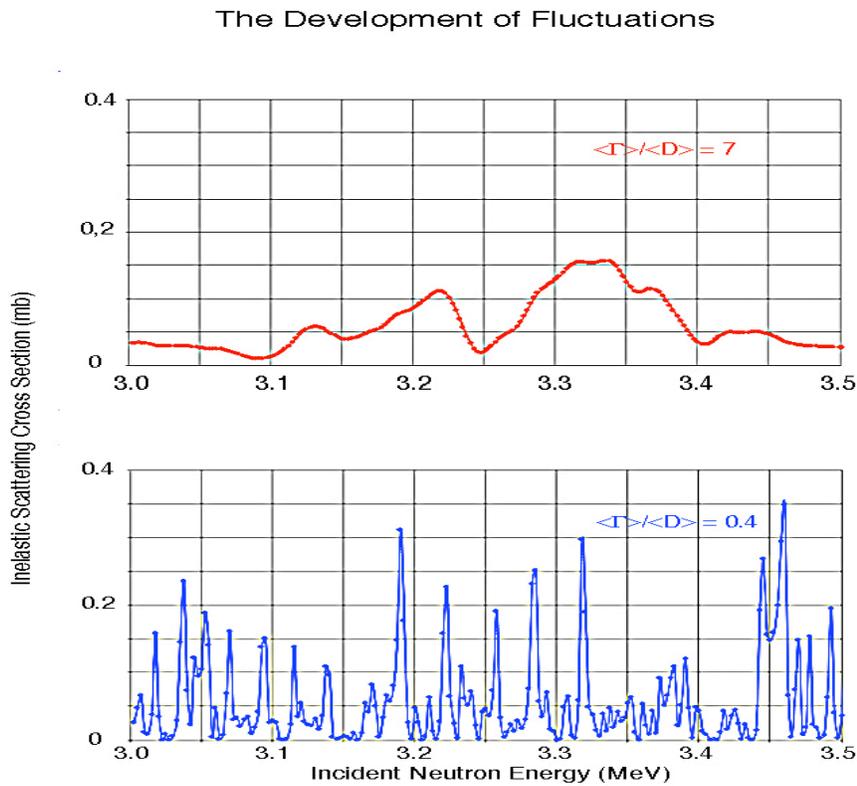

Figure 3. *The inelastic neutron scattering cross section to a state at 1 MeV for the two values $<\Gamma>/<D>$ = 0.4 and 7. The underlying fine-structure resonance energies are the same in the two cases.*

The spacings are chosen randomly from the same Wigner distribution, using the same set of random numbers, and the neutron widths for elastic and inelastic scattering are chosen randomly from Porter–Thomas distributions. For both values of the strength function, a constant, average value for the sum of the eliminated channels is assumed. The onset of fluctuations is clearly seen.

## 4. AN R–FUNCTION MODEL OF FLUCTUATING ECONOMIC INDICES

The R-function theory of Thomas, described and used above to study the onset of fluctuations in nuclear reactions is adapted to model the variations observed in the Dow Jones Industrial Average on a typical trading day. Values of the fundamental parameters of the theory that best describe the three basic forms of the daily DJIA are obtained. The predictive features of the theory are discussed.

It is well known that the fluctuations observed in the time-dependent DJIA have a nearest-neighbor spacing distribution that is of a Wigner form associated with random matrices belonging to a Gaussian Orthogonal Ensemble (Plerou *et al.*, 2000). Observed departures from the Wigner distribution were interpreted as a demonstration of long-term correlations in the market process. In the present analysis, a Wigner spacing distribution of adjacent states in the DJIA assumed. The widths of the states are selected from $\chi^2$ – distributions, with degrees of freedom that depend on the number of modes of trading associated with a given state. The importance of high-resolution, time-dependent information in the analysis of an economic index is necessary in determining the basic parameter (average width/average spacing) for the fine structure states. A minimum time-resolution of 10 seconds is required if a reliable value of <Γ>/<D> is to be obtained from the data. Examples of the three structural modes, calculated using an R-function model, are shown in the following Figure.

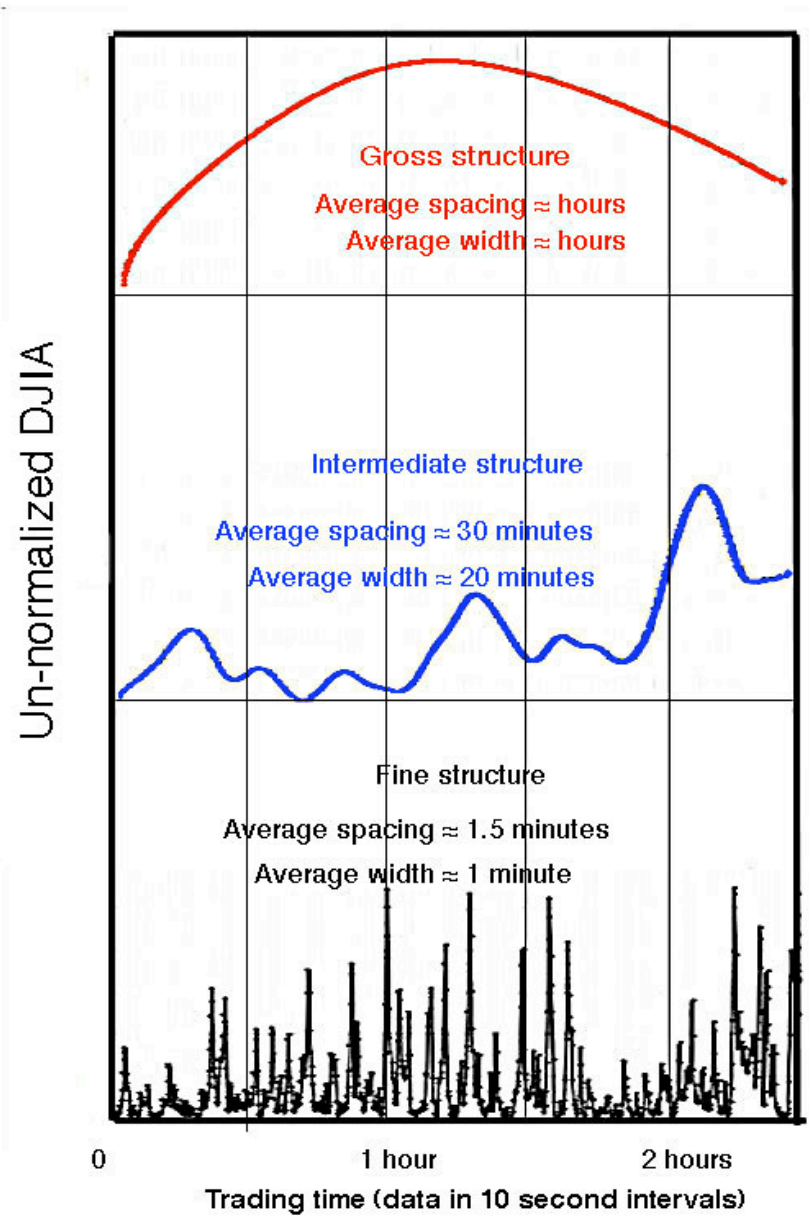

Figure 4. *The three basic components of the DJIA calculated using the R-function theory of Thomas (1955). The key parameter in each case is the value of <Γ>/<D>. The underlying states are uncorrelated.*

An example of the combined fine and intermediate structure for values of <Γ>/D> = 0.4 and 4 is shown in Figure 5.

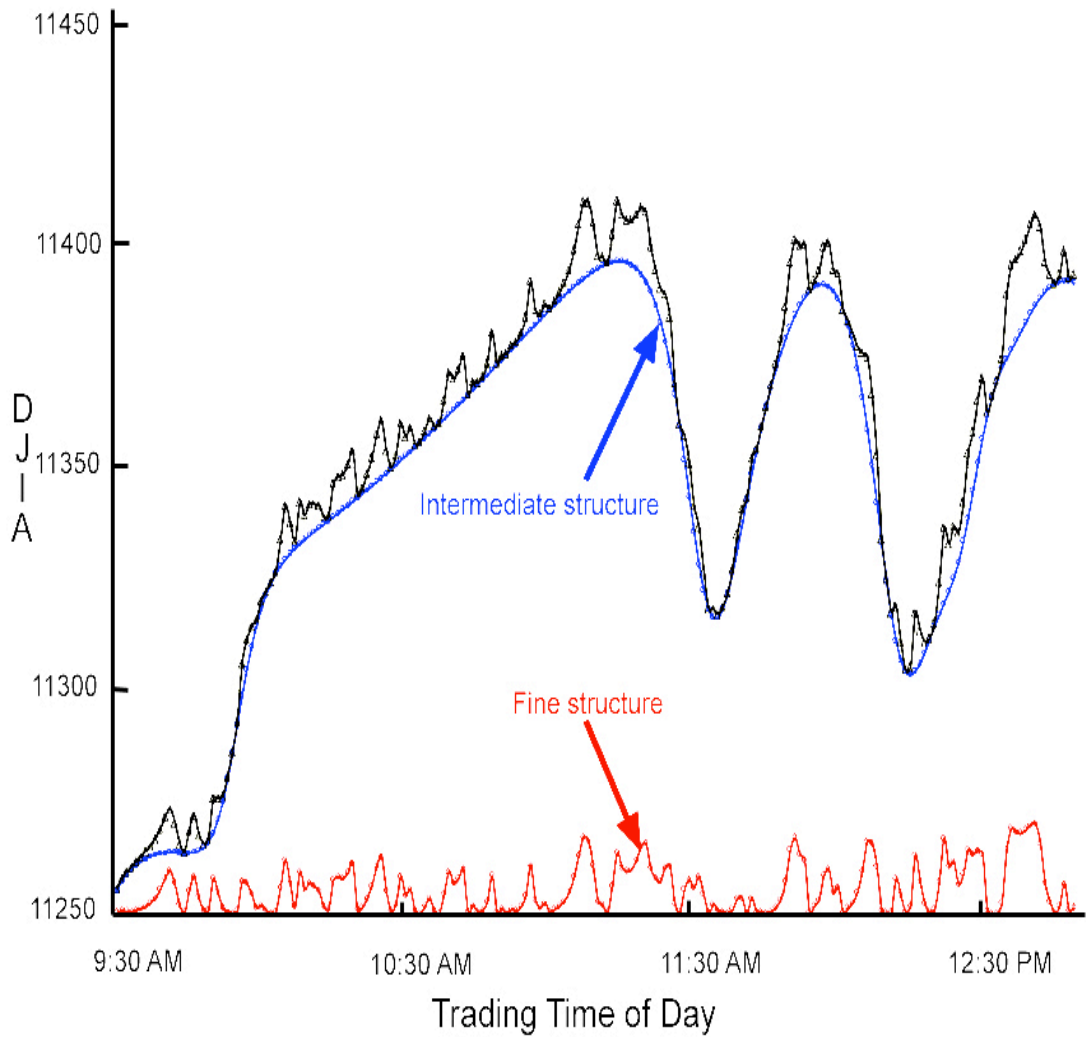

Figure 5. *The calculated fine and intermediate structures are combined to give a form typical of that observed on a given trading day. Here, the average spacing of the fine structure is chosen to be 4 minutes, and the average width is 1.6 minutes. The underlying states of the two structures are assumed to be uncorrelated. The rich forms of the fine structure states, characteristic of the R-function model, are clearly seen.*

The fine structure observed can have an average spacing as small as 1 minute, and an average width of 30 seconds. The value of the fine structure strength function depends upon the average load, bandwidth, processing capacity, and delays in buy and sell orders (the *latency*) associated with High-Frequency Trading on a given day. It is assumed that the underlying states of the intermediate structure are not correlated with the states of the fine structure; this is not necessarily the case.

It is proposed that the values of $<\Gamma>/<D>$ for the fine *and* intermediate structure of the states of the DJIA, on any day, be obtained by analyzing the data in the first one- to two-hours of trading, and that these values be used as *predictors* of the values for the remaining trading time on the given day. This procedure assumes that there is sufficient *inertia* in the fine and intermediate structure associated with trades on the given day; there is ample numerical evidence in support of this assumption.

Sophisticated non-linear fitting programs, based on R-matrix theory, are available in Nuclear and Atomic Physics, and recently in Astrophysics. Precise resonance parameters have been obtained from measured cross sections, and their values compared with theoretical predictions.

**5. CONCLUSIONS**

The R-function theory of Thomas, a practical variant of R-matrix theory, was developed to analyze and understand nuclear and atomic reaction cross-sections. Here, it used to account, quantitatively, for the three major structures observed in a typical economic index, such as the DJIA. A fundamental parameter in the theory is the "strength function", <Γ>/<D>. The method emphasizes not only the *spacing* between adjacent fluctuations but also the *widths*, or *lifetimes* of the fluctuations. Knowledge of the lifetimes of the three major components of an economic index is critically important in developing any successful investment strategy.

**References**

Bohr A and Mottelson B R 1975 *Nuclear Structure* Vol. II (Reading Mass: Benjamin)

Ericson T E O 1960 *Phys. Rev. Lett*. **5** 430

———— 1963 *Ann. Phys*. (N.Y.) **23** 390

Feshbach H 1962 *Ann. Phys*. (N.Y.) **19**, 287

Firk F W K, Lynn J E and Moxon M C 1963 *Proc. Phys. Soc.* **82** 477

Firk F W K 2010 *arXiv*:**1003.2002v2**

Lane A M and Thomas R G 1958 *Rev. Mod. Phys.* **30** 257

Lynn J E 1968 *The Theory of Neutron Resonance Reactions* (Oxford: The Clarendon Press)

Moldauer P A 1964a *Phys. Rev*. **135** 624B

———— 1964b *ibid* **136** 949B

Plerou V, Gopikrishnan P, Rosenow B, Amaral L A N, and Stanley H E 2000 *Physica* **A 287** 374

Porter C E and Thomas R G 1956 *Phys. Rev.* **104** 483

Reich C W and Moore M S 1958 *Phys. Rev.* **111** 929

Thomas R G 1955 *Phys. Rev.* **97** 224

Vogt E 1958 *Phys. Rev.* **112** 203

———1960 *ibid.* **118** 724

Wigner E P and Eisenbud L 1947 *Phys. Rev.* **72** 29

Wigner E P 1957 *Proc. Conf. on Neutron Physics by Time-of-Flight Methods (Gatlinburg TN, Nov. 1956)* Oak Ridge National Lab. Report. ORNL-2309

Wu C-P, Firk F W K and Phillips T W 1970 *Nucl. Phys.* **A147** 19